\documentclass[twocolumn,showpacs,preprintnumbers,amsmath,amssymb,prl]{revtex4-2}

\usepackage{amsmath,amssymb,amsfonts}
\usepackage{graphicx}
\usepackage{dcolumn}
\usepackage{bm}
\usepackage{hyperref}
\usepackage{booktabs}
\usepackage{tikz}
\usetikzlibrary{arrows.meta, positioning, shapes.geometric, calc}
\usepackage{braket}

\newcommand{\Mhat}{\hat{M}}
\newcommand{\Hhat}{\hat{H}}

\newcommand{\FEPS}{\textsc{feps}}

\newcommand{\sh}{\sinh}
\newcommand{\ch}{\cosh}
\newcommand{\EqTF}{\mathbb{E}_{\rm TF}}
\newcommand{\EqQS}{\mathbb{E}_{\rm QS}^{\rm(reg)}}
\newcommand{\eps}{\epsilon}

\begin{document}

\title{Complementary Quantum Time Distributions from a Single Operational Protocol}

\author{Mathieu Beau}
\affiliation{Department of Physics, University of Massachusetts Boston,
Boston, Massachusetts 02125, USA}

\date{\today}

\begin{abstract}
A single operational protocol based on free evolution and projective measurements yields inequivalent quantum time distributions through distinct post-processing procedures. We construct an activity-based time-of-flow (TF) distribution and a presence-based quantum stroboscopic (QS) distribution, providing complementary operational notions of time. 
{Applied to tunneling through a rectangular barrier, the regional QS mean saturates as the barrier width increases, whereas the TF mean first decreases and eventually grows for larger widths. These contrasting behaviors provide an operational interpretation of the Hartman effect, the apparent saturation of tunneling delay with increasing barrier width, in terms of quantum time distributions associated with flow through the exit region and occupation within the barrier, capturing the roles of early penetration, dominant reflection, and spectrally filtered transmission.}
\end{abstract}

%\pacs{03.65.Ta, 03.65.Ca, 03.67.-a}
\keywords{time in quantum mechanics; time-of-flow; time-energy uncertainty;
tunneling time; quantum stroboscopy}

\maketitle

%% =====================================================================
\textit{Introduction.}---How do we measure time in quantum physics?  
Unlike position or momentum, time is not represented by a canonical self-adjoint operator~\cite{Pauli1933}, and typically enters as an external parameter. Assigning a \emph{time of occurrence} to events such as detections, state transitions, or barrier traversal therefore requires an operational construction rather than a direct observable.

A wide range of approaches has been developed to address this problem \cite{Muga2008,Muga2009}. Relational and quantum-clock frameworks describe time as emerging from correlations between systems~\cite{PageWootters1983,Rovelli1996,Giovannetti2015,Maccone2020}. 
Other approaches introduce time observables through POVMs or restricted operator constructions, including Kijowski-type distributions, and generalized measurement formalisms~\cite{Kijowski1974,Werner1986,Grot1996,Muga2000,Galapon2002,Muga2008}. 
Operational models based on detectors, weak measurements, or repeated interrogation define time through measurement protocols, often at the cost of backaction or model dependence~\cite{Allcock1969a,Allcock1969b,Allcock1969c,Brunetti2002,Wiseman2002,Ramos2020}. 
Taken together, these approaches indicate that time in quantum mechanics is not accessed as an intrinsic observable, but reconstructed from measurement statistics. Temporal quantities are thus inherently contextual, depending on the observable, the protocol, and the post-processing of measurement records.

Recent works have introduced an explicitly operational framework based on repeated projective measurements on identically prepared systems~\cite{Beau2025a,Beau2025b,Lloyd2026,Martellini2026}. 
Within this setting, two distinct time distributions have emerged. 
The time-of-flow (TF) distribution reconstructs transition events~\cite{Beau2025a} and recovers flux-based arrival-time results~\cite{Beau2025b}, while the quantum stroboscopic (QS) distribution reconstructs temporal occupation from repeated measurements~\cite{Lloyd2026}. 
Although both rely on the same free-evolution projective sampling (\FEPS{}) protocol, they {yield} two inequivalent time distributions.

In this work, we show that inequivalent quantum time distributions can arise from distinct post-processing procedures applied to a single operational protocol. This leads to two complementary notions of time: an activity-based time-of-flow (TF) distribution~\cite{Beau2025a,Beau2025b} and an occupation-based quantum stroboscopic (QS) distribution~\cite{Lloyd2026}. These distributions do not represent alternative approximations of a single quantity, but correspond to distinct operational questions about the dynamics. {Their mathematical inequivalence follows directly from the distinct post-processings,
$\pi_{\mathrm{TF}}\propto|\dot p|$ and $\pi_{\mathrm{QS}}\propto p$.
Ref.~\cite{Lloyd2026} had already noted that the flux-based quantum-flow distribution
introduced in Ref.~\cite{Beau2024b} can be reconstructed from position-resolved
stroboscopic data. Here, we consider the broader TF construction introduced in
Refs.~\cite{Beau2025a,Beau2025b}, applicable to both discrete and continuous systems,
and show through its operational comparison with the QS distribution that the two
constructions answer distinct questions and yield physically informative contrasts in
concrete settings.}

We illustrate this distinction with the Rabi model and apply it to tunneling dynamics, where the contrasting behaviors of TF and QS yield an operational interpretation of the Hartman effect~\cite{Hartman1962,Buttiker1983,Hauge1989,Steinberg1995,Winful2006} in terms of flow through the exit region and occupation within the barrier.

\begin{figure*}[t]
\centering
\begin{tikzpicture}[
  every node/.style={font=\small},
  protobox/.style={draw, rounded corners=6pt,
    draw=gray!60!black, fill=gray!11,
    minimum width=7.0cm, minimum height=1.1cm,
    align=center, inner sep=7pt},
  tfell/.style={draw, ellipse, draw=teal!70!black, fill=teal!12,
    minimum width=3.6cm, minimum height=1.3cm,
    align=center, inner sep=5pt, font=\small\bfseries},
  qsell/.style={draw, ellipse, draw=violet!65!black, fill=violet!9,
    minimum width=3.6cm, minimum height=1.3cm,
    align=center, inner sep=5pt, font=\small\bfseries},
  qflow/.style={draw, diamond, aspect=2.4, draw=teal!60!black, fill=teal!6,
    minimum width=3.8cm, minimum height=1.0cm, align=center, inner sep=4pt},
  qclock/.style={draw, diamond, aspect=2.4, draw=violet!55!black, fill=violet!5,
    minimum width=3.8cm, minimum height=1.0cm, align=center, inner sep=4pt},
  arr/.style={->, >=stealth, thick, draw=gray!60},
  arrteal/.style={->, >=stealth, thick, draw=teal!65!black},
  arrviolet/.style={->, >=stealth, thick, draw=violet!60!black},
  lbl/.style={font=\scriptsize, fill=white, inner sep=2pt, align=center}]
\draw[dashed, gray!45, thick] (-7.6, 2.75) -- (5.6, 2.75);
\draw[dashed, gray!45, thick] (-7.6, 0.80) -- (5.6, 0.80);
\node[font=\scriptsize\itshape, text=gray!65!black, align=right,
  text width=2.6cm] at (-6.8, 4.40) {Operational layer\\(protocol)};
\node[font=\scriptsize\itshape, text=gray!65!black, align=right,
  text width=2.6cm] at (-6.8, 1.75) {Distributional layer\\(post-protocol)};
\node[font=\scriptsize\itshape, text=gray!65!black, align=right,
  text width=2.9cm] at (-6.9, -0.30) {Conceptual layer\\ ({Interpretation})};
\node[protobox] (FEPS) at (0, 4.40)
  {\textbf{FEPS protocol}\\[3pt]$p(t)=\mathrm{Tr}(\hat{M}\hat{\rho}_t)$};
\node[tfell] (TF) at (-3.2, 1.75)
  {TF\\[2pt]\normalfont\scriptsize activity-based};
\node[qsell] (QS) at (3.2, 1.75)
  {QS\\[2pt]\normalfont\scriptsize presence-based};
\node[qflow]  (QF) at (-3.2, -0.30) { \hspace{0.005cm}  Transition Time \hspace{0.005cm}  };
\node[qclock] (QC) at ( 3.2, -0.30) {Occupancy Time};
\draw[arr] (FEPS.south) -- ++(0,-0.50) -| (TF.north)
  node[lbl, pos=0.55, xshift=-30pt]
  {$\pi_{\rm TF}\!\propto\!\left|\dfrac{d}{dt}p(t)\right|$};
\draw[arr] (FEPS.south) -- ++(0,-0.50) -| (QS.north)
  node[lbl, pos=0.55, xshift=25pt]{$\pi_{\rm QS}\propto p(t)$};
\draw[arrteal] (TF.south) -- (QF.north)
  node[lbl, midway, xshift=34pt]{{$\hat{J}=\tfrac{i}{\hbar}[\hat{H},\hat{M}]$}};
\draw[arrviolet] (QS.south) -- (QC.north)
  node[lbl, midway, xshift=15pt]{$\hat{M}$};
\end{tikzpicture}
\caption{Three-layer structure of operational quantum timing.
\textit{Operational layer}: the \FEPS{} protocol reconstructs
$p(t)=\mathrm{Tr}(\hat{M}\hat{\rho}_t)$ from single-shot projective
measurements on independent copies, avoiding quantum Zeno effects.
\textit{Distributional layer}: two inequivalent post-processings yield
an activity-based TF distribution~\cite{Beau2025a,Beau2025b}
($\pi_{\rm TF}\propto|\dot{p}|$)
and a presence-based QS distribution~\cite{Lloyd2026}
($\pi_{\rm QS}\propto p$).
\textit{Interpretive layer}: TF is associated with the generator of change
${\hat{J}=\tfrac{i}{\hbar}[\hat{H},\hat{M}]}$, while QS corresponds to the observable
$\hat{M}$ itself.
In position measurements, TF reduces to the probability current $|j(x,t)|$,
whereas QS yields the density $\rho(x,t)$.
In settings where measurements project onto specific quantum states,
QS coincides with quantum stroboscopic time distributions~\cite{Lloyd2026}.}
\label{fig:landscape}
\end{figure*}

%% =====================================================================

\textit{Protocol and the two distributions.}---
We consider an operational protocol in which identically prepared copies of the system evolve freely, and each copy is measured once at a chosen time. Repeating the experiment for different measurement times allows one to reconstruct
\[
p(t)=\mathrm{Tr}[\hat{M}\hat{\rho}_t],
\]
where $\hat{M}$ is a projector and $\hat \rho_t$ is the density operator. 
This \textit{free-evolution projective sampling} (\FEPS{}) protocol was introduced to reconstruct the time of flow (TF) distribution for discrete systems~\cite{Beau2025a}, extended to continuous systems~\cite{Beau2025b}, and further applied to quantum stroboscopic (QS) settings~\cite{Lloyd2026}. By relying on single-shot measurements on independent copies rather than continuous monitoring, it avoids quantum Zeno effects.

From the same signal $p(t)$, two inequivalent time distributions follow from distinct post-processing:
\begin{equation}
\pi_{\rm TF}(t) = \frac{1}{Z_{\rm TF}}|\dot{p}(t)|,\qquad
\pi_{\rm QS}(t) = \frac{1}{Z_{\rm QS}}p(t),
\label{eq:TF_QS}
\end{equation}
with $Z_{\rm TF}=\int_{t_i}^{t_f}|\dot{p}(t)|\,dt$ and $Z_{\rm QS}=\int_{t_i}^{t_f} p(t)\,dt$.
In settings where measurements project onto specific quantum states, the QS distribution coincides with the time distributions reconstructed from quantum stroboscopic measurements~\cite{Lloyd2026}.
The three-layer structure is summarized in Fig.~\ref{fig:landscape}, where the interpretive layer will be clarified below.

%=====================================================================
\textit{Rabi illustration and interpretation.}---To illustrate the distinction between the two distributions, we consider the Rabi model
$\Hhat=(\hbar\omega_0/2)\hat{\sigma}_x$, with initial state $\ket{0}$ and target state $\ket{1}$, using $\hat{M}=\ket{1}\bra{1}$. 
The \FEPS{} protocol reconstructs $p_1(t)=\sin^2(\omega_0 t/2)$ over one period $T_R=2\pi/\omega_0$.

The TF distribution resolves the non-monotonic dynamics by conditioning on the sign of $\dot{p}_1$, distinguishing time of arrival (TOA) when $\dot{p}_1>0$ from time of departure (TOD) when $\dot{p}_1<0$. 
Restricting to the arrival half-cycle $[0,\pi/\omega_0]$ yields
\[
\pi^{\rm TOA}(t)=\frac{\omega_0}{2}\sin(\omega_0 t),
\]
with mean $T_R/4$, corresponding to maximal transition activity; the second half-cycle analogously defines a TOD distribution.

In contrast, the QS distribution normalizes $p_1(t)$ over the full period,
\[
\pi_{\rm QS}(t)=\frac{\omega_0}{\pi}\sin^2(\omega_0 t/2),
\]
yielding a mean $T_R/2$ at maximal occupation. 
Unlike TF, QS does not distinguish arrival from departure and assigns equal weight to both.

The difference between $T_R/4$ and $T_R/2$ reflects two inequivalent interpretations: TF measures the time of change (activity), while QS measures the time of presence (occupation). 
This distinction is particularly clear in the stationary limit $[\hat{H},\hat{\rho}]=0$, where TF vanishes (no events) whereas QS reduces to a uniform distribution over $[0,t_f]$, reflecting persistent occupation.

\textit{Matter-wave dynamics.}---For $\Hhat=\hat{p}^2/(2m)+V(\hat{x})$ and
$\hat{M}=\Mhat_{x_0}=\int_{x_0}^\infty|x'\rangle\langle x'|dx'$, 
the \FEPS{} protocol reconstructs the cumulative distribution
$F(x_0,t)=\mathrm{Tr}(\hat M_{x_0}\hat \rho_t)=\int_{x_0}^\infty\rho(x,t)\,dx$, where $\rho(x,t)$ is the density function. 
The TF distribution is obtained from its time derivative,
\begin{equation}
\pi_{\rm TF}(t)=|\partial_t F(x_0,t)|/Z_{\rm TF}=|j(x_0,t)|/Z_{\rm TF},
\label{eq:TF_Fx}
\end{equation}
using $\partial_t\Mhat_{x_0}=(i/\hbar)[\Hhat,\Mhat_{x_0}]$, which defines the current operator $\hat{J}(x_0)$~\cite{Beau2025a}. The TF distribution reproduces the flux-based expressions for the time-of-arrival obtained across different frameworks, including standard flux formulations, Bohmian trajectory approaches, and decoherent histories \cite{Leavens1993,Leavens1998,Muga2000,Halliwell2009,Beau2024a,Beau2024b}.
{Owing to the modulus and to the state-dependent normalization $Z_{\rm TF}$, $\pi_{\rm TF}$ is not linear in $\hat\rho$ and therefore does not define a POVM. This is fully consistent with the no-go result of Vona \emph{et al.}~\cite{Vona2013}, who proved that, in the presence of quantum backflow, no POVM can reproduce flux-based (Bohmian) arrival-time statistics. The present construction falls outside the assumptions of this theorem rather than contradicting it: linearity holds at the level of the measured signal $F(x_0,t)$, each run involves a single projective measurement, while the time distribution is obtained by classical post-processing of the ensemble record and thus need not be of POVM form. For flux-positive states, $\pi_{\rm TF}$ coincides with the normalized Bohmian arrival-time distribution for crossing trajectories~\cite{Vona2013,Leavens1998}; more generally, conditioning on the sign of $\dot p$ separates arrival (TOA) from departure (TOD) events, so that backflow appears as an operationally resolved feature rather than an inconsistency~\cite{Beau2024b}.}
{Recent debate over Bohmian first-arrival times has focused on the relation between trajectory-defined quantities and operational measurement statistics. Das and D\"urr proposed testing first-arrival distributions defined by Bohmian trajectories in the absence of detectors~\cite{DasDurr2019}. Goldstein, Tumulka, and Zangh\`i~\cite{Goldstein2024a,Goldstein2024b} emphasized that arrival without a detector, arrival in its presence, and detector-click time need not coincide, and that the outcome statistics of a fixed experimental procedure are governed by a POVM. Drezet~\cite{Drezet2024} argued that the Das-D\"urr distribution could nevertheless be reconstructed in principle using weakly coupled detectors and a nonlinear, trajectory-informed post-selection procedure. In the present framework, no trajectory reconstruction is required: $\pi_{\rm TF}$ is obtained by classical post-processing of the family $F(x_0,t)$, reconstructed from single-shot projective measurements at selected times on independent copies, each evolving unmonitored until sampling, thereby avoiding cumulative measurement backaction.}

For the regional operator $\hat{M}_{[x_0,x_0+L]}=\int_{x_0}^{x_0+L}|x\rangle\langle x|dx$, the protocol yields
\begin{equation}
\pi_{\rm QS}^{\rm (reg)}(t)=p_{[x_0,x_0+L]}(t)/Z_{\rm QS}^{\rm (reg)},
\end{equation}
where $p_{[x_0,x_0+L]}(t) = \int_{x_0}^{x_0+L} \rho(x,t)dx=F(x_0,t)-F(x_0+L,t)$.
Taking the limit $L\to0$ gives the local QS distribution
\begin{equation}
\pi_{\rm QS}^{\rm (loc)}(t)=-\partial_x F(x_0,t)/Z_{\rm QS}^{\rm (loc)}=\rho(x_0,t)/Z_{\rm QS}^{\rm (loc)},
\label{eq:QS_Fx}
\end{equation}
which can be implemented via a finite-difference approximation of $\partial_x\Mhat_{x_0}$. This construction is equivalent to the quantum stroboscopic protocol \cite{Lloyd2026}. 

Thus, TF and QS correspond to time and space derivatives of the same cumulative distribution $F(x_0,t)$.
This establishes two inequivalent, yet complementary distributions, extracted from the same protocol: TF encodes directional flow, including quantum backflow~\cite{Bracken1994,Beau2024b}, whereas QS encodes local occupation and coincides with quantum clock distributions~\cite{Maccone2020,Lloyd2026}.
{More generally, since $\hat M_{[x_0,x_0+L]}=\hat M_{x_0}-\hat M_{x_0+L}$, every regional signal is determined by the measured one-parameter family $\{F(x_0,t)\}$. In the tunneling analysis below, the TF and QS distributions are accordingly obtained from different members of this same family, namely $F(L,t)$ for the exit flow and $F(0,t)-F(L,t)$ for the barrier occupation, rather than from literally identical signals.}

\begin{figure}[h!]
\centering
\includegraphics[width=\columnwidth]{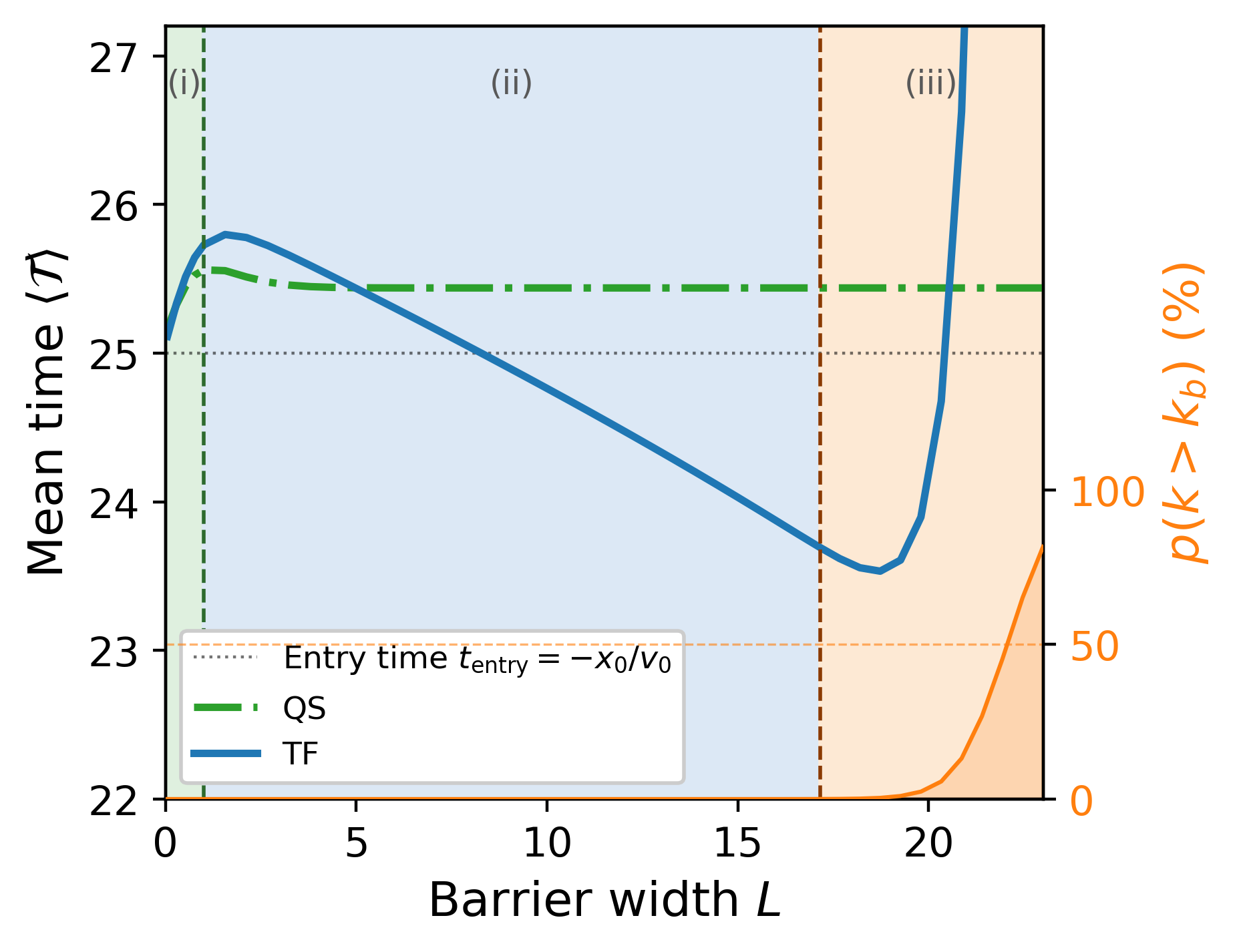}
\caption{
\textbf{
Mean TF (solid) and QS (dash-dotted) times versus barrier width $L${, obtained from the same measured family of cumulative signals}.}
QS saturates with increasing $L$, while after an initial decrease in the Hartman regime, TF {eventually exhibits a rapid growth with $L$}, driven by spectral filtering and the increasing contribution of above-barrier components. The secondary axis shows the fraction of transmitted above-barrier components $p(k>k_b)$, indicating a crossover in the transmitted spectrum. {The shaded regions (i)--(iii) and vertical dashed lines indicate the regimes $L \lesssim 1/\varepsilon(k_0)$, $1/\varepsilon(k_0) \lesssim L \lesssim L^\ast$, and $L \gtrsim L^\ast$, where $L^\ast$ is the leading-order estimate of the above-barrier crossover.
Results are shown for $m=0.5$, $\hbar=1$, $V_0=2$, $k_0=1.0$, $\sigma_k=0.05$, $x_0=-50$.}
}\label{fig:tunneling}
\end{figure}

% -----------------------------------------------------------------------
\textit{Tunneling time: operational interpretation of the Hartman effect.}—
We consider a Gaussian wave packet 
$\phi(k) =(2\pi\sigma_k^2)^{-1/4} e^{-(k-k_0)^2/(4\sigma_k^2)}e^{-ikx_0}$ 
incident on a rectangular barrier of height $V_0$ and width $L$, with central energy $E_0<V_0$, so that the incident wave packet is predominantly in the tunneling regime, while retaining an exponentially small above-barrier component. 

To probe the transition time through the barrier, we realize the \FEPS{} protocol using the projector $\hat{M}_L=\int_L^{\infty} |x\rangle\langle x|dx$, which reconstructs the cumulative probability at the barrier exit. 
The corresponding TF distribution measures the time of transmitted flow across $x=L$. 
In the narrow-bandwidth ($\sigma_k \ll k_0$) and far-field ($|x_0|\gg\sigma_x$) limits, its mean takes the spectral form~\cite{Wigner1955,Hauge1989,Beau2025a,Beau2025b}
\begin{equation}
\langle \mathcal{T} \rangle_{\rm TF}
= \mathbb{E}_{\rm TF}\!\left[
\frac{-x_0}{v(k)}+\tau_W^{\rm(exit)}(k,L)
\right],
\label{eq:TF_PRL}
\end{equation}
where $\mathbb{E}_{\rm TF}[f]=(1/Z_{\rm TF})\int dk\,|\phi_k|^2T_k f(k)$,
$T_k=|t(k,L)|^2$, where $t(k,L)$ is the transmission {amplitude} at the exit of the barrier, and $\tau_W^{\rm(exit)}(k,L)=v(k)^{-1}\left(\partial_k\arg[t(k,L)e^{ikL}]\right)$ is the Wigner phase time evaluated at the exit~\cite{Wigner1955,Hauge1989}.

The behavior of this observable is governed by the dominant spectral contributions. In the opaque regime, $1 \ll \epsilon(k_0) L$, and below the crossover to above-barrier components, the transmission coefficient $T_k \sim e^{-2\epsilon(k)L}$ (here $\epsilon(k)\equiv \sqrt{k_b^2-k^2}$ and $k_b^2=2mV_0/\hbar^2$) induces an exponential filtering that shifts the transmitted spectrum toward higher momenta. This results in an effective velocity $v(k_\ast)>v_0$ and a corresponding time advancement relative to free propagation
\begin{equation}
\langle \mathcal T\rangle_{\rm TF}
\approx
-\frac{x_0}{v(k_\ast)}+\tau_W^{\rm (\infty)}(k_\ast),
\label{eq:TF_largeL}
\end{equation}
where $k_\ast$ is the momentum selected by spectral filtering, with $(k_\ast-k_0)/k_0 \approx (2\sigma_k^2/\epsilon(k_0))\,L$, as long as the shift remains small compared to the spectral width. This approximation holds while tunneling contributions dominate, i.e., for $L < L^\ast$, with
$L^\ast = (k_b - k_0)^2/(4\sigma_k^2\,\epsilon(k_0))$ (yielding $L^\ast \sim 17$ for Fig.~\ref{fig:tunneling}). The intermediate regime $\epsilon_0^{-1}\ll L\ll L^\ast$ corresponds to the Hartman plateau, where $\tau_W^{\rm(exit)}(k,L)\to \tau_W^{\rm(\infty)}=2m/(\hbar k\,\epsilon(k))$~\cite{Hartman1962,Hauge1989}. For $L \gg L^\ast$, sub-barrier contributions are exponentially suppressed and the transmitted signal becomes dominated by above-barrier components $k>k_b$~\cite{Muga2002,Winful2006}, leading to a linear growth of the mean TF, as illustrated in Fig.~\ref{fig:tunneling}.

The QS distribution, probing the temporal occupation within the barrier region and constructed from the \FEPS{} protocol with the regional projector $\hat{M}_{[0,L]}=\int_0^L |x\rangle\langle x|dx$, brings a different perspective. 
Its mean can be expressed as a spectral average~\cite{SM}: 
\begin{equation}
\langle \mathcal{T} \rangle_{\rm QS}^{\rm(reg)}
= \mathbb{E}_{\rm QS}^{\rm(reg)}\!\left[
\frac{-x_0}{v(k)}+\tau_W^{\rm(exit)}(k,L)
\right]+\mathcal{R}(L),
\label{eq:meanQS}
\end{equation}
where $\mathbb{E}_{\rm QS}^{\rm(reg)}[f]=(1/Z_{\rm QS}^{\rm(reg)})\int dk\,|\phi_k|^2\tau_D(k,L) f(k)$ and $\tau_D(k,L)$ is the Smith dwell time \cite{Smith1962,Hauge1989}, which saturates when $L$ is large, and where the remainder $\mathcal{R}(L)\approx 0$ when $L$ is large. 
Hence, the mean value of the QS distribution \eqref{eq:meanQS} converges in the opaque limit $\epsilon(k_0) L \gg 1$ to
\begin{equation}
\langle \mathcal{T} \rangle_{\rm QS}^{\rm (reg)} \to \frac{-x_0}{v(k_0)}+\tau_W^{(\infty)}(k_0).
\end{equation}

\textit{Quantum time uncertainty.}—
In the narrow-bandwidth regime $\sigma_k\ll k_0$, the spread of the TF distribution at the barrier exit admits the estimate
\begin{equation}\label{Eq:DeltaTF}
\Delta \mathcal T_{\rm TF}
\approx \frac{1}{2 v(k_\ast)\sigma_k}
\left(
1
-
\frac{L}{2L_{c}}
\right),\; L\ll L_{c} ,
\end{equation}
where the crossover scale $L_{c} = \epsilon(k_0)^3/(2\,k_b^2\,\sigma_k^2)$ 
marks the breakdown of the sub-barrier Gaussian saddle approximation and the onset of the crossover toward barrier-top and eventually over-barrier contributions.
The expression \eqref{Eq:DeltaTF} reflects the spread of the TF distribution in terms of the transmitted activity, with $v(k_\ast)=\hbar k_\ast/m$. 
{Its origin is the following: in the sub-barrier saddle regime, the exponential filter $e^{-2\epsilon(k)L}$ both selects the faster momentum $k_\ast$ and, being log-convex in $k$ ($\partial_k^2\epsilon<0$), slightly \emph{increases} the effective bandwidth $\sigma_{\rm eff}(L)\gtrsim\sigma_k$ entering the stationary-phase estimate; both effects reduce the temporal spread, yielding the mild linear decrease in Eq.~\eqref{Eq:DeltaTF}. Numerics confirm that this estimate is tight for $L\lesssim L^\ast$ (see the plots of the distributions and of their spreads in~\cite{SM}). Beyond $L^\ast$, the saddle description breaks down: the transmitted spectrum is truncated near the barrier top and reduces to the narrow above-barrier tail of the incident packet, while the residual sub-barrier peak coexists with a broad, resonantly modulated above-barrier contribution. In this crossover regime, the numerically computed spread grows rapidly with $L$ and lies well above Eq.~\eqref{Eq:DeltaTF}, which tracks only the sub-barrier saddle component and should then be regarded as a lower estimate rather than a controlled approximation~\cite{SM}.}

For the regional QS distribution, one finds in the same narrow-band regime \cite{SM}
\begin{equation}\label{Eq:DeltaQS}
\Delta \mathcal T_{\rm QS}^{\rm(reg)}
\approx
\frac{1}{2\;\sigma_k\,v(k_0)},    
\end{equation}
which remains independent of $L$ in the opaque limit. 
This qualitative difference reflects the nature of the observables: TF probes transmitted activity at the exit and therefore depends on the selected momentum $k_\ast$, whereas QS captures in-barrier occupation and remains controlled by the incident spectral width.

\textit{Discussion.}— The QS and TF constructions answer two distinct operational questions arising from the \FEPS{} measurement protocol through different post-processing procedures, {applied, in the tunneling case, to different members of the same measured family $F(x_0,t)$}: the QS distribution asks \emph{when is probability present within the barrier}, while the TF distribution asks \emph{when does probability flow across the exit}. 

From the presence-based (QS) perspective, the occupation inside the barrier is dominated by the leading tail of the wave packet, which penetrates the barrier before the arrival time of the incident peak $t_{\rm entry}\approx -x_0/v_0$. 
This early contribution is followed by a persistent but weak presence from the fraction of the wave packet that has penetrated the barrier. However, since the dominant part of the incident amplitude is reflected once the packet reaches the barrier ({in the opaque regime,} $j(0,t)<0$ shortly after the entry time $t_{\rm entry}$, see \cite{SM}), there is no sustained injection of probability into the barrier, so the occupation does not grow at later times and remains concentrated around the entry time, up to corrections associated with the Wigner phase time. As a result, the peak time of residence remains close to the entry time and does not drift with increasing barrier width, leading to saturation. 

From the flow-based (TF) perspective, once probability has penetrated the barrier, it contributes to a transmitted current at the exit, whose temporal profile is again centered near the entry time, up to corrections, as a result of the concentration of the QS distribution around $t_{\rm entry}$. These corrections include a Wigner phase time contribution associated with penetration through the barrier, as well as an advancement arising from spectral filtering, which enhances higher-momentum components arriving earlier than the incident peak $t_{\rm entry}^\ast\sim -x_0/v(k_\ast)<t_{\rm entry}$. 

Within this framework, the Hartman effect does not reflect a well-defined traversal duration, but rather emerges from the combined effect of early penetration, dominant reflection, and spectrally filtered transmission, captured through two complementary temporal distributions associated with presence and flow.
{This physical picture; saturating phase and dwell times, dominant reflection, and spectral filtering of the transmitted components; largely coincides with the interpretations previously reached through group-delay and phase-time analyses of the Hartman effect~\cite{Hartman1962,Hauge1989,Winful2006}. The added value of the present framework lies not in a new mechanism, but in its operational underpinning: these interpretations are recovered as statements about two explicitly measurable time distributions obtained by different post-processings of a single projective-sampling scheme.}

\textit{Conclusion.}— We have shown that quantum timing does not correspond to a unique probability distribution, but emerges from distinct post-processing procedures applied to a single operational protocol. Two inequivalent distributions arise, providing different perspectives on the dynamics: an activity-based time-of-flow (TF), associated with probability current, and an occupation-based quantum stroboscopic (QS) distribution, associated with probability density and integrated presence.

More specifically, TF characterizes when probability flows across a boundary defined by the observable through the generator ${\hat{J}=\tfrac{i}{\hbar}[\hat{H},\hat{M}]}$, while QS characterizes when it is present within the subspace associated with the projector $\hat{M}$. This distinction applies to both discrete and continuous systems and depends only on the choice of measurement operator. In the Rabi model, QS describes the temporal occupation of the target state, while TF resolves the timing of transition events between the initial and target states, illustrating the general interpretation of QS as presence and TF as activity.

Applied to tunneling, this complementarity provides an operational interpretation of the Hartman effect. The saturation of the QS mean reflects the limited occupation within the barrier, while the TF distribution captures the transmission dynamics, including early contributions from spectrally selected higher-momentum components and the eventual crossover to above-barrier propagation.

More broadly, this framework demonstrates that a single operational protocol can give rise to multiple inequivalent time distributions depending on how the measurement record is analyzed, highlighting the operational and contextual character of time in quantum theory.

\bibliographystyle{apsrev4-2}
\bibliography{Refv2}

%\clearpage

\newpage

\onecolumngrid

\vspace{5mm} %5mm vertical space

\begin{center}
\textbf{\large SUPPLEMENTARY MATERIAL}
\end{center}

\tableofcontents

\section{Tunneling-Time Derivations}
\label{sec:tunneling}
%% ============================================================

%% ----------------------------------------------------------
\subsection{Setup: scattering states and exact formulas}
\label{sec:setup}
%% ----------------------------------------------------------

We consider a rectangular barrier of height $V_0$ and width $L$ on $[0,L]$.
For a stationary state with wavenumber $k>0$ and energy
$E_k=\hbar^2k^2/(2m)<V_0$, define
\[
\epsilon(k)=\sqrt{k_b^2-k^2},
\qquad
k_b=\frac{\sqrt{2mV_0}}{\hbar}.
\]
{Throughout this Supplemental Material we write $\eps_0\equiv\epsilon(k_0)=\sqrt{k_b^2-k_0^2}$; the barrier-top wavenumber is always denoted $k_b$.}
Inside the barrier, the scattering solution reads
\begin{equation}
\psi_k(x)=t(k,L)e^{ikL}
\left[
\cosh\!\big(\epsilon(x-L)\big)
+i\frac{k}{\epsilon}\sinh\!\big(\epsilon(x-L)\big)
\right],
\qquad 0\le x\le L,
\label{eq:psi_inside}
\end{equation}
where the transmission amplitude is
\begin{equation}
t(k,L)=e^{-ikL}\,
\frac{-2i\epsilon k}
{(\epsilon^2-k^2)\sinh(\epsilon L)-2ik\epsilon\cosh(\epsilon L)},
\label{eq:tk}
\end{equation}
so that
\begin{equation}
T(k,L)\equiv |t(k,L)|^2
=
\frac{4\epsilon^2 k^2}
{(\epsilon^2-k^2)^2\sinh^2(\epsilon L)+4k^2\epsilon^2\cosh^2(\epsilon L)}.
\label{eq:Tk}
\end{equation}
The full wavepacket evolves as
\begin{equation}
\Psi(x,t)=\int_0^\infty dk\,\phi(k)\psi_k(x)e^{-iE_k t/\hbar},
\end{equation}
with
\[
|\phi(k)|^2=\frac{1}{\sqrt{2\pi}\sigma_k}
\exp\!\left[-\frac{(k-k_0)^2}{2\sigma_k^2}\right],
\qquad
\sigma_k\ll k_0,
\]
and $\phi(k)=|\phi(k)|e^{-ikx_0}$, where $x_0<0$ is the initial
packet center.

\paragraph{Smith dwell time.}
Using Eqs.~\eqref{eq:psi_inside} and \eqref{eq:Tk}, the Smith dwell time
\[
\tau_D(k,L)\equiv \frac{1}{v(k)}\int_0^L |\psi_k(x)|^2\,dx\quad,
\qquad
v(k)=\frac{\hbar k}{m},
\]
takes the explicit form
\begin{equation}
\tau_D(k,L)
=
\frac{mk}{\hbar\epsilon}\,
\frac{
2\epsilon(\epsilon^2-k^2)L+(\epsilon^2+k^2)\sinh(2\epsilon L)
}{
(\epsilon^2-k^2)^2\sinh^2(\epsilon L)+4k^2\epsilon^2\cosh^2(\epsilon L)
}.
\label{eq:tauD_exact}
\end{equation}
In the opaque-barrier limit $\epsilon(k)L\gg1$, this saturates to \cite{Smith1962}
\begin{equation}
\boxed{
\tau_D^{(\infty)}(k)
=
\frac{2mk}{\hbar\,\epsilon(k)\,[k^2+\epsilon(k)^2]} .
}
\label{eq:tauD_inf_k}
\end{equation}

\paragraph{Wigner phase time.}
The transmission amplitude can be written as
$t(k,L)=|t(k,L)|e^{i\phi_k}$ with
\[
\phi_k=-kL+\tilde\phi_k,
\]
where
\begin{equation}
\cos\tilde\phi_k=\frac{2\eps k\ch(\eps L)}{D_k^{1/2}},
\qquad
\sin\tilde\phi_k=-\frac{(\eps^2-k^2)\sh(\eps L)}{D_k^{1/2}},
\label{eq:phi_decomp}
\end{equation}
and
\begin{equation}\label{Eq:Dk}
D_k=(\eps^2-k^2)^2\sh^2(\eps L)+4\eps^2k^2\ch^2(\eps L).    
\end{equation}
Equivalently,
\begin{equation}
\tilde\phi_k
=
\arctan\!\left(
-\frac{\eps^2-k^2}{2k\eps}\tanh(\eps L)
\right).
\end{equation}
The Wigner phase time at the barrier exit is defined from the phase
of the transmitted wave at $x=L$ \cite{Wigner1955}, namely
\begin{equation}
\tau_W^{\rm exit}(k,L)=\frac{1}{v(k)}\,\partial_k\tilde\phi_k
=\hbar\,\partial_E\tilde\phi_k,
\qquad
v(k)=\frac{\hbar k}{m},
\end{equation}
which yields the exact expression 
\begin{equation}
\tau_W^{\rm exit}(k,L)
=
\frac{m}{\hbar k\eps}\,
\frac{
(k^2+\eps^2)^2\,\sh(2\eps L)
+2\eps k^2(\eps^2-k^2)L
}{
(k^2+\eps^2)^2\sh^2(\eps L)+4k^2\eps^2
}.
\label{eq:tauW_exact}
\end{equation}
In the opaque-barrier limit $\eps L\gg1$, this saturates to
\begin{equation}
\boxed{
\tau_W^{(\infty)}(k)=\frac{2m}{\hbar k\,\eps(k)},
}
\end{equation}
which is the Hartman limit.

%% ----------------------------------------------------------
\subsection{TF mean time: exact derivation}
\label{sec:SM_TF}
%% ----------------------------------------------------------
\paragraph{Transmitted current in the narrow-bandwidth and far-field regimes.}
The transmitted current at \(x=L\) is
\begin{align}
J(L,t)&=\frac{\hbar}{2mi}
\bigl[\Psi^*(x,t)\partial_x\Psi(x,t)-\Psi(x,t)\partial_x\Psi^*(x,t)\bigr]_{x=L}\\
&={\frac{\hbar}{2m}\int_{-\infty}^{+\infty}\!dk\int_{-\infty}^{+\infty}\!dk'\,(k+k')\,\phi_k\,\phi^*_{k'}\,t(k,L)\,t^*(k',L)\,e^{i(k-k')L}\,e^{-i(E_k-E_{k'}) t/\hbar}}\\
&={\frac{\hbar}{2m}\int_{-\infty}^{+\infty}\!dk\int_{-\infty}^{+\infty}\!dk'\,(k+k')\sqrt{T(k,L)\,T(k',L)}\,\sqrt{g_k\, g_{k'}}\;e^{i[\theta_k(L)-\theta_{k'}(L)]}\,e^{-i(E_k-E_{k'}) t/\hbar}} ,
\end{align}
{where $g_k=|\phi_k|^2$ and where we used the transmitted wave at the exit, $\psi_k(L)=t(k,L)e^{ikL}=|t(k,L)|\,e^{i\tilde\phi_k}$, so that the total phase is}
$$
\theta_k(L) = -kx_0+\tilde{\phi}_k .
$$

For the calculations below, we consider the following assumptions:
\begin{itemize}
    \item \textbf{Narrow-bandwidth regime}: \(\sigma_k\ll k_0\), with \(k_0>0\).
    \item \textbf{Far-field regime}: \(|x_0|\gg \sigma_x\), where \(\sigma_x=1/(2\sigma_k)\), and $x_0<0$.
\end{itemize}
Under these assumptions, the initial Gaussian wavepacket is well localized far from the barrier at \(t=0\), so that its penetration into the barrier is negligible initially. Moreover, for a right-moving packet sharply peaked around \(k_0>0\), backflow effects are negligible, and one may approximate \(J(L,t)\ge 0\){; for such effectively flux-positive dynamics the TF distribution coincides with the Bohmian arrival-time distribution, and the non-POVM character of flux-based arrival statistics established by Vona \emph{et al.}~\cite{Vona2013} plays no quantitative role at the exit (see the discussion in the main text)}.

Hence, the mean TF is approximated as
\begin{equation}
\langle \mathcal{T}\rangle_{\rm TF}
\equiv
\dfrac{\int_0^{+\infty}dt\, t\, J(L,t)}{\int_0^{+\infty}dt\, J(L,t)}
\approx
\frac{1}{Z_{\rm TF}}\int_{-\infty}^{+\infty}dt\, t\, J(L,t),
\label{eq:MeanTF}
\end{equation}
where $Z_{\rm TF} \equiv \int_{-\infty}^{+\infty}dt\, J(L,t)$
since the contribution from \(t<0\) is negligible in the far-field regime.

\paragraph{Time-integrated exit flux.}

Using the identity
$$
\int_\mathbb{R} e^{-i(E_k-E_{k'})t/\hbar}dt
=2\pi\hbar\delta(E_k-E_{k'})=\dfrac{2\pi m}{\hbar k}\delta(k-k') ,
$$
we find the expression for the normalization factor of the TF distribution:
\begin{equation}
Z_{\rm TF}\equiv\int_0^\infty J(L,t)\,dt
=2\pi\int_0^\infty dk\,|\phi_k|^2\,T(k,L) .
\label{eq:ZTF}
\end{equation}

\paragraph{First temporal moment.}
Using the identity
$$
\int_\mathbb{R} t\,e^{-i(E_k-E_{k'})t/\hbar}dt
=2\pi i\hbar\,\delta'(E_k-E_{k'}) ,
$$ 
and
$$
\delta'(E_k-E_{k'})=\dfrac{m}{\hbar^2 k}\delta'(k-k')-
\dfrac{m^2}{\hbar^4 k^2}\delta(k-k') ,
$$
the second (odd) term vanishes since $v(-k)^3=-v(k)^3$
while $T_{-k}=T_k$, $|\phi_{-k}|^2=|\phi_k|^2$.
Integrating the $\delta'(k-k')$ term by parts:
\begin{equation}
\int_0^\infty t\,J(L,t)\,dt
=2\pi\int_0^\infty dk\,|\phi_k|^2\,T(k,L)
\!\left(\frac{-x_0}{v(k)}+\tau_W(k,L)\right).
\label{eq:NTF}
\end{equation}

\paragraph{Expected value of the TF distribution.}
Dividing Eq.~\eqref{eq:NTF} by Eq.~\eqref{eq:ZTF}:
\begin{equation}
\boxed{
\langle\mathcal{T}\rangle_{\rm TF}
=\EqTF\!\left[\frac{-x_0}{v(k)}+\tau_W(k,L)\right],
}
\label{eq:TF_mean}
\end{equation}
where $\EqTF[f]=(1/Z_{\rm TF})\int dk\,|\phi_k|^2 T(k,L)f(k)$.

\paragraph{Large-$\eps_0 L$ asymptotics.}
In the regime $\eps_0 L\gg 1$, from equation \eqref{Eq:Dk}, we find the asymptote
$$
D_k\approx \dfrac{1}{4}(\eps^2+k^2)^2 e^{2\eps L} ,
$$
and from \eqref{eq:Tk} 
$$
T(k,L)\approx \dfrac{16\eps^2 k^2}{(\eps^2+k^2)^2} e^{-2\eps L} .
$$
Hence, the exponential weight 
$$T(k,L)\,|\phi_k|^2\propto
\exp{\left[-2\eps(k)L-\dfrac{(k-k_0)^2}{2\sigma_k^2}\right]}$$
is maximized at
\begin{equation}
k^*=k_0+\delta k,\qquad
\delta k\simeq\frac{2\sigma_k^2 k_0}{\eps_0}\,L,
\end{equation}
shifting the transmitted spectrum toward higher momenta.
Since $\tau_W(k,L)\to\tau_W^{(\infty)}(k)=2m/(\hbar k\eps(k))$ saturates,
\begin{equation}
\langle\mathcal{T}\rangle_{\rm TF}
\approx\frac{-x_0}{v(k^*)}+\tau_W^{(\infty)}(k^*),
\end{equation}
which decreases linearly in $L$ below the Hartman plateau, as shown in Figure 2 in the main body of the paper.

\paragraph{Very-large-$L$ asymptotics}

For $L\gg L^\ast$, the sub-barrier contribution is exponentially suppressed by the factor
$e^{-2\varepsilon(k)L}$, so that the transmission-weighted integral is eventually dominated
by momenta $k>k_b$, i.e. by energies above the barrier height $V_0$. In this regime, the
wavefunction in the barrier region is oscillatory rather than evanescent (one has to change the hyperbolic functions in \eqref{eq:psi_inside} by the trigonometric analog), and the transmission
coefficient is no longer exponentially small in $L$ (although it may retain an oscillatory,
resonant dependence on $L$). As a result, the mean TF time recovers an approximately linear
growth,
$$
\langle t\rangle_{\rm TF}\approx \frac{-x_0}{v_{\rm eff}}+\tau_W^{\rm exit}(k_{\rm eff},L),
$$
where $v_{\rm eff}=\hbar k_{\rm eff}/m>v_0$ is associated with the effective momentum
selected by the transmission-weighted above-barrier contribution, and where the exit
phase time now grows approximately linearly with the barrier width,
$\tau_W^{\rm exit}\approx L/v'_{\rm eff}$, with
$v'_{\rm eff}=\hbar\sqrt{k_{\rm eff}^2-k_b^2}/m$ the group velocity inside the barrier
region for above-barrier propagation. The deviation from a
simple classical extrapolation is then governed both by this spectral reshaping and
by the oscillatory, resonant structure that $\tau_W^{\rm exit}$ can display.

The crossover length at which the above-barrier fraction of the transmitted flux
reaches $50\%$ is, to leading order,
\begin{equation}
    L^\ast = \frac{({k_b}-k_0)^2}{4\sigma_k^2\,\varepsilon(k_0)},
    \qquad \varepsilon(k) \equiv \sqrt{{k_b^2}-k^2},
    \label{eq:Lstar}
\end{equation}
obtained by equating the dominant exponents of the two contributions,
$e^{-2\varepsilon(k_0)L^\ast}=e^{-({k_b}-k_0)^2/2\sigma_k^2}$,
and is valid when $({k_b}-k_0)\gg\sigma_k$.
{(Being a leading-exponent estimate, $L^\ast$ slightly underestimates the numerically observed $50\%$ crossover; for the parameters of the main text one finds $L^\ast\simeq 17$, while the crossover of the transmitted weight occurs at $L\simeq 22$, see Fig.~2 of the main text.)}

\subsection{Local QS mean}

The local QS distribution at $x=L$ uses $\hat{M}=|L\rangle\langle L|$,
giving $P_L(t)=\rho_t(L)$ and spectral weight $T(k,L)/v(k)$:
\begin{equation}
\langle\mathcal{T}\rangle_{\rm QS}^{\rm(loc)}
=\mathbb{E}_{\rm QS}^{\rm(loc)}\!\left[\frac{-x_0}{v(k)}
+\tau_W(k,L)\right],
\label{eq:QS_loc_mean}
\end{equation}
where $\mathbb{E}_{\rm QS}^{\rm(loc)}[f]=
(1/Z_{\rm QS}^{\rm(loc)})\int dk\,|\phi_k|^2(T_k/v(k))f(k)$
with $Z_{\rm QS}^{\rm(loc)}=\int dk\,|\phi_k|^2 T_k/v(k)$.
The local QS mean differs from the TF mean only in its spectral
weight ($T_k/v$ vs $T_k$), producing $O(\sigma_k^2)$ corrections
in the narrow-bandwidth regime.

%% ----------------------------------------------------------
\subsection{Regional QS mean time}
\label{sec:SM_QS}
%% ----------------------------------------------------------

\paragraph{Quasi-stationary partition function.}
For $\hat{M}_L=\int_0^L|x\rangle\langle x|dx$, the integrated
occupation is $Z_{\rm QS}^{\rm(reg)}=\int_0^\infty P_L(t)\,dt$
with $P_L(t)=\int_0^L|\Psi(x,t)|^2 dx$.
Using the time integral identity and defining $\tau_D(k)$, we find
\begin{equation}
Z_{\rm QS}^{\rm(reg)}\approx
2\pi\int_\mathbb{R}dk\,|\phi_k|^2\,\tau_D(k).
\label{eq:ZQS}
\end{equation}
For $\eps_0 L\gg 1$, $\tau_D(k_0)\to\tau_D^{(\infty)}(k_0)$
({Eq.~\eqref{eq:tauD_inf_k}}), so $Z_{\rm QS}^{\rm(reg)}\to
2\pi\,\tau_D^{(\infty)}(k_0)$.

\paragraph{First temporal moment $N_{\rm QS}$.}
The first moment 
$$
N_{\rm QS}=\int_0^\infty t\,P_L(t)\,dt\approx \int_{-\infty}^\infty t\,P_L(t)\,dt
$$ 
follows
the same $\delta'$ expansion as we used to derive the spectral form of the mean of the TF distribution.
The odd-parity term vanishes.
The remaining contribution splits as
\begin{align}
N_{\rm QS}=\pi\int_\mathbb{R} dk\,|\phi_k|^2
\biggl[
&\underbrace{-\frac{x_0}{v(k)}\,\tau_D(k)}_{\text{entry time}}
+\underbrace{\frac{\partial_k\phi_k}{v(k)^2}\,
\frac{\hbar}{m}\int_0^L|\psi_k|^2dx}_{\tau_W\tau_D/v}
+\underbrace{\frac{\hbar^2 I_k}{m^{2}v(k)^2}}_{\text{phase integral}}
\biggr],
\end{align}
where $I_k=\int_0^L(\partial_k X_k(x))|\psi_k(x)|^2 dx$
and $X_k(x)$ is the accumulated spatial phase of $\psi_k(x)$
inside the barrier.

\paragraph{Evaluation of $I_k$.}
For the rectangular barrier, from Eq.~\eqref{eq:psi_inside}:
$\partial_k X_k(x)=(k^2/\eps^2)(x-L)/S_k(x)^2$
with $|\psi_k(x)|^2=T_k S_k(x)^2$. Hence
\begin{equation}
I_k=\frac{k^2}{\eps^2}\,T(k,L)\int_0^L(x-L)\,dx
=-\frac{k^2}{2\eps^2}\,T(k,L)\,L^2
=-\frac{m^2 v(k)^2}{2\hbar^2\eps^2}\,T(k,L)\,L^2.
\label{eq:Ik}
\end{equation}
Since $T(k,L)\sim e^{-2\eps L}$, the contribution of $I_k$
constitutes the remainder
\begin{equation}
\mathcal{R}(L)
=\frac{\pi}{Z_{\rm QS}^{\rm(reg)}}
\int dk\,|\phi_k|^2\,T(k,L)\,\frac{m^2 L^2}{2\hbar^2\eps^2}
=O(L^2 e^{-2\eps_0 L})\to 0.
\label{eq:RL}
\end{equation}

\paragraph{Mean of the QS distribution.}
Collecting entry-time and phase-time contributions and dividing
by $Z_{\rm QS}^{\rm(reg)}$:
\begin{equation}
\boxed{
\langle\mathcal{T}\rangle_{\rm QS}^{\rm(reg)}
=\EqQS\!\left[\frac{-x_0}{v(k)}+\tau_W(k,L)\right]
+\mathcal{R}(L),
}
\label{eq:QS_mean}
\end{equation}
where $\EqQS[f]=(1/Z_{\rm QS}^{\rm(reg)})
\int dk\,|\phi_k|^2\tau_D(k,L)f(k)$.
This is Eq.~({7}) of the main text.

\textbf{Saturation.} For $\eps_0 L\gg 1$,
$\tau_W(k,L)\to\tau_W^{(\infty)}(k_0)$ saturates{, while the spectral weight
becomes $L$-independent and remains sharply concentrated around $k_0$ in the
narrow-band limit}, so
\begin{equation}
\langle\mathcal{T}\rangle_{\rm QS}^{\rm(reg)}
\;\to\;\frac{-x_0}{v(k_0)}+\tau_W^{(\infty)}(k_0),
\end{equation}
in contrast to the TF mean which decreases linearly in the Hartman regime and grows in the $L\gg L^\ast$ regime.

%% ============================================================
\section{Quantum time uncertainty}
\label{sec:TimeUncertainty}
%% ============================================================

\subsection{Unified stationary-phase estimates for the mean and spread of TF and regional QS}

In this section we derive simple narrow-band estimates for both the mean time and the temporal spread of the TF and regional QS distributions. 
The common idea is that, once the relevant signal is written in spectral form, its center is determined by the stationary phase condition, while its width is set by the Gaussian spectral envelope around the stationary point.

\paragraph{General principle.}
Consider a signal of the form
\begin{equation}
S(t)\propto \int dk\, g(k)\,e^{i[\varphi(k)-E_k t/\hbar]},
\label{eq:generic_signal}
\end{equation}
where $g(k)$ is sharply peaked around some momentum $k_\ast$ with effective width $\sigma_{\rm eff}$. 
Expanding
\begin{equation}
\varphi(k)-\frac{E_k t}{\hbar}
\approx
\varphi(k_\ast)-\frac{E_{k_\ast}t}{\hbar}
+(k-k_\ast)\big[\varphi'(k_\ast)-v_\ast t\big]
+\cdots,
\end{equation}
with
\[
v_\ast=\frac{1}{\hbar}\left.\frac{dE}{dk}\right|_{k_\ast}=\frac{\hbar k_\ast}{m},
\]
the stationary-phase condition yields the center
\begin{equation}
t_{\rm c}=\frac{\varphi'(k_\ast)}{v_\ast},
\label{eq:generic_center}
\end{equation}
while a Gaussian envelope $g(k)\sim e^{-(k-k_\ast)^2/(4\sigma_{\rm eff}^2)}$ leads to
\begin{equation}
S(t)\propto e^{-\sigma_{\rm eff}^2 v_\ast^2 (t-t_{\rm c})^2},
\qquad
|S(t)|^2\propto e^{-2\sigma_{\rm eff}^2 v_\ast^2 (t-t_{\rm c})^2},
\end{equation}
hence the temporal spread
\begin{equation}
\Delta t \approx \frac{1}{2\sigma_{\rm eff}v_\ast}.
\label{eq:generic_width}
\end{equation}
We now apply this principle to TF and regional QS.

\subsection{Gaussian estimate of the TF mean time and temporal spread}
\label{app:TF_gaussian_derivation}

In this section we derive the Gaussian approximation used in the main text for the
mean time and temporal spread of the TF distribution at the barrier exit.
The starting point is the transmitted wavefunction evaluated at $x=L$,
\begin{equation}
\Psi_T(L,t)=\frac{1}{\sqrt{2\pi}}\int dk\,\phi(k)\,t(k,L)\,e^{ikL}e^{-iE_k t/\hbar},
\qquad
E_k=\frac{\hbar^2k^2}{2m}.
\label{eq:psi_exit_SM}
\end{equation}
For a Gaussian incident packet centered at $k_0$,
\begin{equation}
\phi(k)=A\exp\!\left[-\frac{(k-k_0)^2}{4\sigma_k^2}\right]e^{-ikx_0},
\label{eq:phi_gauss_SM}
\end{equation}
with $x_0<0$ and $\sigma_k\ll k_0$.

We write the transmission amplitude in polar form as
\begin{equation}
t(k,L)e^{ikL}=|t(k,L)|\,e^{i\varphi_{\rm TF}(k,L)},
\label{eq:t_polar_SM}
\end{equation}
where
\begin{equation}
\varphi_{\rm TF}(k,L):=\arg\!\big[t(k,L)e^{ikL}\big]
\label{eq:varphiTF_def_SM}
\end{equation}
is the phase of the transmitted wave at the barrier exit. Equation~\eqref{eq:psi_exit_SM}
then becomes
\begin{equation}
\Psi_T(L,t)=\frac{A}{\sqrt{2\pi}}\int dk\,
\exp\!\left[-\frac{(k-k_0)^2}{4\sigma_k^2}\right]
|t(k,L)|\,
e^{\,i\Phi_{\rm TF}(k,t)},
\label{eq:psi_exit_phase_SM}
\end{equation}
with
\begin{equation}
\Phi_{\rm TF}(k,t):=-kx_0+\varphi_{\rm TF}(k,L)-\frac{E_k t}{\hbar}.
\label{eq:PhiTF_def_SM}
\end{equation}

\paragraph{Dominant momentum and effective width.}
The modulus of the transmitted integrand is governed by the spectral weight
\begin{equation}
F_{\rm TF}(k,L):=|\phi(k)|^2\,T(k,L),
\qquad
T(k,L)=|t(k,L)|^2.
\label{eq:FTF_def_SM}
\end{equation}
In the narrow-band regime we assume that $F_{\rm TF}(k,L)$ has a single dominant maximum
at $k_\ast=k_\ast(L)$, defined by
\begin{equation}
\left.\partial_k \ln F_{\rm TF}(k,L)\right|_{k=k_\ast}=0.
\label{eq:kstar_def_SM}
\end{equation}
For the Gaussian packet~\eqref{eq:phi_gauss_SM}, this condition gives
\begin{equation}
k_\ast-k_0=\sigma_k^2\,\partial_k\ln T(k_\ast,L),
\label{eq:kstar_shift_SM}
\end{equation}
which makes explicit the barrier-induced spectral filtering.

Expanding $\ln F_{\rm TF}(k,L)$ to second order around $k_\ast$ yields
\begin{equation}
\ln F_{\rm TF}(k,L)\approx
\ln F_{\rm TF}(k_\ast,L)
-\frac{(k-k_\ast)^2}{2[\sigma_{\rm eff}^{\rm(TF)}(L)]^2},
\label{eq:lnFTF_expand_SM}
\end{equation}
where the effective transmitted width is defined by
\begin{equation}
\frac{1}{[\sigma_{\rm eff}^{\rm(TF)}(L)]^2}
=
-\left.\partial_k^2\ln F_{\rm TF}(k,L)\right|_{k=k_\ast}
=
\frac{1}{\sigma_k^2}
-
\left.\partial_k^2\ln T(k,L)\right|_{k=k_\ast}.
\label{eq:sigmaeff_TF_SM}
\end{equation}
Equivalently, at the amplitude level one may write
\begin{equation}
|\phi(k)t(k,L)|
\approx
|\phi(k_\ast)t(k_\ast,L)|
\exp\!\left[-\frac{(k-k_\ast)^2}{4[\sigma_{\rm eff}^{\rm(TF)}(L)]^2}\right].
\label{eq:amp_gaussian_SM}
\end{equation}

\paragraph{Quadratic expansion of the phase.}
We now expand the phase \eqref{eq:PhiTF_def_SM} to second order around $k_\ast$.
Setting
\[
q:=k-k_\ast,
\]
we obtain
\begin{equation}
\Phi_{\rm TF}(k,t)\approx
\Phi_\ast(t)
+q\,A(t)
+\frac{q^2}{2}\,B(t),
\label{eq:phase_quad_SM}
\end{equation}
with
\begin{align}
\Phi_\ast(t)
&=
-k_\ast x_0+\varphi_{\rm TF}(k_\ast,L)-\frac{E_{k_\ast}t}{\hbar},
\\[1mm]
A(t)
&=
-x_0+\partial_k\varphi_{\rm TF}(k_\ast,L)-v_\ast t,
\label{eq:AofT_SM}
\\[1mm]
B(t)
&=
\partial_k^2\varphi_{\rm TF}(k_\ast,L)-\frac{\hbar t}{m},
\label{eq:BofT_SM}
\end{align}
where
\begin{equation}
v_\ast=\frac{1}{\hbar}\left.\frac{dE_k}{dk}\right|_{k_\ast}
=\frac{\hbar k_\ast}{m}.
\label{eq:vstar_SM}
\end{equation}

Using Eqs.~\eqref{eq:amp_gaussian_SM} and \eqref{eq:phase_quad_SM}, the transmitted wavefunction
takes the Gaussian-integral form
\begin{equation}
\Psi_T(L,t)\approx
\mathcal N\,
e^{i\Phi_\ast(t)}
\int dq\,
\exp\!\left[
-\frac{q^2}{4[\sigma_{\rm eff}^{\rm(TF)}]^2}
+iA(t)q+\frac{i}{2}B(t)q^2
\right],
\label{eq:psi_gaussian_integral_SM}
\end{equation}
where $\mathcal N$ is a slowly varying prefactor independent of $q$ at this order.

\paragraph{Exact Gaussian integration.}
Introduce the complex coefficient
\begin{equation}
\alpha(t):=
\frac{1}{4[\sigma_{\rm eff}^{\rm(TF)}]^2}
-\frac{i}{2}B(t).
\label{eq:alpha_def_SM}
\end{equation}
Then Eq.~\eqref{eq:psi_gaussian_integral_SM} becomes
\begin{equation}
\Psi_T(L,t)\approx
\mathcal N\,e^{i\Phi_\ast(t)}
\int dq\,e^{-\alpha(t)q^2+iA(t)q}.
\label{eq:psi_alpha_SM}
\end{equation}
Provided $\Re\alpha>0$, the Gaussian integral yields
\begin{equation}
\int dq\,e^{-\alpha q^2+iAq}
=
\sqrt{\frac{\pi}{\alpha}}\,
\exp\!\left[-\frac{A^2}{4\alpha}\right].
\label{eq:gaussian_integral_SM}
\end{equation}
Hence
\begin{equation}
\Psi_T(L,t)\approx
\mathcal N\,e^{i\Phi_\ast(t)}
\sqrt{\frac{\pi}{\alpha(t)}}
\exp\!\left[-\frac{A(t)^2}{4\alpha(t)}\right].
\label{eq:psi_after_integral_SM}
\end{equation}

Taking the modulus squared gives
\begin{equation}
|\Psi_T(L,t)|^2
\approx
|\mathcal N|^2\frac{\pi}{|\alpha(t)|}
\exp\!\left[
-\frac{A(t)^2}{4\alpha(t)}
-\frac{A(t)^2}{4\alpha^\ast(t)}
\right].
\label{eq:psi_sq_prelim_SM}
\end{equation}
Using
\[
\frac{1}{4\alpha}+\frac{1}{4\alpha^\ast}
=\frac{\Re(\alpha)}{2|\alpha|^2},
\qquad
\Re(\alpha)=\frac{1}{4[\sigma_{\rm eff}^{\rm(TF)}]^2},
\]
we obtain
\begin{equation}
|\Psi_T(L,t)|^2
\approx
|\mathcal N|^2\frac{\pi}{|\alpha(t)|}
\exp\!\left[
-\frac{A(t)^2}{8[\sigma_{\rm eff}^{\rm(TF)}]^2|\alpha(t)|^2}
\right].
\label{eq:psi_sq_exactquad_SM}
\end{equation}
Equation~\eqref{eq:psi_sq_exactquad_SM} is the full quadratic approximation. The factor
$|\alpha(t)|^{-1}$ and the dependence of $|\alpha(t)|$ on $t$ encode the weak dispersive
broadening arising from the second derivative of the phase.

\paragraph{Narrow-band limit.}
In the regime relevant to the main text, the quadratic phase correction is small over the
transmitted spectral width,
\begin{equation}
|B(t)|\,(\sigma_{\rm eff}^{\rm(TF)})^2\ll1.
\label{eq:weak_dispersion_condition_SM}
\end{equation}
Then
\begin{equation}
\alpha(t)\approx \frac{1}{4[\sigma_{\rm eff}^{\rm(TF)}]^2},
\qquad
|\alpha(t)|^2\approx \frac{1}{16[\sigma_{\rm eff}^{\rm(TF)}]^4},
\label{eq:alpha_approx_SM}
\end{equation}
and Eq.~\eqref{eq:psi_sq_exactquad_SM} simplifies to
\begin{equation}
|\Psi_T(L,t)|^2
\propto
\exp\!\left[-2[\sigma_{\rm eff}^{\rm(TF)}]^2A(t)^2\right].
\label{eq:psi_sq_narrowband_SM}
\end{equation}
Since
\begin{equation}
A(t)=
-x_0+\partial_k\varphi_{\rm TF}(k_\ast,L)-v_\ast t
=
-v_\ast\left(t-t_{\rm TF}\right),
\label{eq:A_rewrite_SM}
\end{equation}
with
\begin{equation}
t_{\rm TF}=
\frac{-x_0+\partial_k\varphi_{\rm TF}(k_\ast,L)}{v_\ast}
=
-\frac{x_0}{v_\ast}+\tau_W^{\rm exit}(k_\ast,L),
\label{eq:tTF_SM}
\end{equation}
we arrive at
\begin{equation}
|\Psi_T(L,t)|^2
\propto
\exp\!\left[
-2[\sigma_{\rm eff}^{\rm(TF)}]^2v_\ast^2(t-t_{\rm TF})^2
\right].
\label{eq:psi_sq_final_SM}
\end{equation}

\paragraph{Relation to the TF distribution.}
For an outgoing narrow-band transmitted packet at the exit, the probability current satisfies
\begin{equation}
j(L,t)=\frac{\hbar}{m}\Im\!\big[\Psi_T^\ast(L,t)\,\partial_x\Psi_T(L,t)\big]
\approx
v(k_\ast)\,|\Psi_T(L,t)|^2.
\label{eq:j_approx_SM}
\end{equation}
Since the TF distribution is reconstructed from the positive outgoing current,
it inherits the same Gaussian temporal profile,
\begin{equation}
\pi_{\rm TF}(t)\propto
\exp\!\left[
-2[\sigma_{\rm eff}^{\rm(TF)}]^2v(k_\ast)^2(t-t_{\rm TF})^2
\right].
\label{eq:piTF_gaussian_SM}
\end{equation}
Therefore,
\begin{equation}
\boxed{
\Delta\mathcal T_{\rm TF}
\approx
\frac{1}{2\,\sigma_{\rm eff}^{\rm(TF)}(L)\,v(k_\ast)}.
}
\label{eq:DeltaTF_final_SM}
\end{equation}

\paragraph{Leading-order form without spectral filtering.}
If the barrier-induced reshaping is weak, one may set $k_\ast\simeq k_0$ and
$\sigma_{\rm eff}^{\rm(TF)}\simeq \sigma_k$, so that
\begin{equation}
t_{\rm TF}\approx -\frac{x_0}{v_0}+\tau_W^{\rm exit}(k_0,L),
\qquad
v_0=\frac{\hbar k_0}{m},
\label{eq:tTF_simple_SM}
\end{equation}
and
\begin{equation}
\pi_{\rm TF}(t)\propto
\exp\!\left[-2\sigma_k^2v_0^2(t-t_{\rm TF})^2\right],
\qquad
\Delta\mathcal T_{\rm TF}\approx \frac{1}{2\sigma_k v_0}.
\label{eq:TF_simple_SM}
\end{equation}
These expressions correspond to the simplest narrow-band estimate quoted in the main text.

\paragraph{Pre-opaque regime: $\varepsilon_0L\ll1$.}
If the barrier does not yet significantly reshape the spectrum, then
\[
k_\ast\approx k_0,
\qquad
\sigma_{\rm eff}^{\rm(TF)}(L)\approx \sigma_k,
\]
so that
\begin{equation}
\boxed{
\langle \mathcal T\rangle_{\rm TF}
\approx
-\frac{x_0}{v(k_0)}+\tau_W(k_0,L),
\qquad
\Delta \mathcal T_{\rm TF}
\approx
\frac{1}{2\sigma_k v(k_0)}.
}
\label{eq:TF_smallL}
\end{equation}

\paragraph{Opaque regime: $\varepsilon_0L\gg1$.}
For tunneling through a rectangular barrier,
\[
|t(k,L)|\sim \mathcal A(k)\,e^{-\varepsilon(k)L},
\qquad
\varepsilon(k)=\sqrt{\frac{2mV_0}{\hbar^2}-k^2},
\]
so the transmitted spectrum is shifted toward the top of the barrier{; within the Gaussian saddle approximation, the curvature-defined width of the transmitted spectrum is in fact \emph{slightly increased}, $\sigma_{\rm eff}^{\rm(TF)}(L)\gtrsim \sigma_k$ [see Eq.~\eqref{eq:sigma_eff} below], because the filter $e^{-\varepsilon(k)L}$ is log-convex in $k$ ($\partial_k^2\varepsilon<0$)}. 
Thus the relevant momentum becomes $k_\ast(L)$.
Equation~\eqref{eq:DeltaTF_final_SM} remains valid, but now with an $L$-dependent width:
\begin{equation}
\boxed{
\langle \mathcal T\rangle_{\rm TF}
\approx
-\frac{x_0}{v(k_\ast)}+\tau_W(k_\ast,L),
\qquad
\Delta \mathcal T_{\rm TF}
\approx
\frac{1}{2\,\sigma_{\rm eff}^{\rm(TF)}(L)\,v(k_\ast)}.
}
\label{eq:TF_largeL}
\end{equation}
{In particular, within its domain of validity the TF spread \emph{decreases} mildly with $L$, through both the increase of $v(k_\ast)$ and the increase of $\sigma_{\rm eff}^{\rm(TF)}(L)$; the growth of the spread at larger $L$ arises only beyond the saddle regime, as discussed below.}

\paragraph{Effective spectral width in the opaque regime.}
In the opaque tunneling regime, with $\varepsilon(k_0)L\gg1$, the transmitted spectral amplitude may be written, up to slowly varying algebraic prefactors, as
\[
g_{\rm TF}(k;L)\propto \phi(k)\,t(k,L)
\sim
\exp\!\left[-\frac{(k-k_0)^2}{4\sigma_k^2}\right] e^{-\varepsilon(k)L},
\]
where
\[
\varepsilon(k)=\sqrt{k_b^2-k^2},
\qquad
k_b^2=\frac{2mV_0}{\hbar^2}.
\]
Expanding $\varepsilon(k)$ to second order around the dominant momentum $k_\ast$,
\[
\varepsilon(k)\approx \varepsilon_\ast+\varepsilon'_\ast (k-k_\ast)+\frac12 \varepsilon''_\ast (k-k_\ast)^2,
\]
with
\[
\varepsilon''_\ast=-\frac{k_b^2}{\varepsilon_\ast^3},
\]
the combined envelope remains Gaussian,
\[
g_{\rm TF}(k;L)\sim \exp\!\left[-\frac{(k-k_\ast)^2}{4\sigma_{\rm eff}^2(L)}\right],
\]
where
\begin{equation}
\boxed{
\frac{1}{\sigma_{\rm eff}^2(L)}
=
\frac{1}{\sigma_k^2}
-
\frac{2k_b^2 L}{\varepsilon(k_\ast)^3}.
}
\label{eq:sigma_eff}
\end{equation}
This expression is valid as long as the quadratic expansion remains controlled and
$1/\sigma_{\rm eff}^2(L)>0$, i.e. before the sub-barrier saddle approaches the barrier top.
Substituting into {Eq.~\eqref{eq:DeltaTF_final_SM}} yields
\begin{equation}
\boxed{
\Delta \mathcal T_{\rm TF}
\approx
\frac{1}{2 v(k_\ast)}
\left(
\frac{1}{\sigma_k^2}
-
\frac{2k_b^2 L}{\varepsilon(k_\ast)^3}
\right)^{1/2}.
}
\end{equation}

The characteristic scale at which the quadratic curvature vanishes is
\[
L_c = \frac{\varepsilon(k_0)^3}{2k_b^2\sigma_k^2}.
\]
Using $\varepsilon(k_\ast)=\varepsilon(k_0)+O(\sigma_k^2)$ in the narrow-band regime gives
\[
\frac{2k_b^2 \sigma_k^2 L}{\varepsilon(k_\ast)^3}
\approx
\frac{L}{L_c}.
\]
Hence, for $L\ll L_c$,
\begin{equation}
\boxed{
\Delta \mathcal T_{\rm TF}
\approx
\frac{1}{2 v(k_\ast)\sigma_k}
\left(
1-\frac{L}{L_c}
\right)^{1/2}
\approx
\frac{1}{2 v(k_\ast)\sigma_k}
\left(
1-\frac{L}{2L_c}
\right).
}
\end{equation}
The scale $L_c$ marks the breakdown of the sub-barrier Gaussian saddle approximation and the onset of the crossover toward barrier-top and eventually over-barrier contributions.

\paragraph{{Beyond the saddle regime: the status of Eq.~(9) of the main text.}}
{Two scales control the validity of the estimate above: the curvature scale $L_c$, at which the Gaussian saddle itself would flatten, and the above-barrier crossover scale $L^\ast$ of Eq.~\eqref{eq:Lstar}, at which transmitted weight becomes dominated by momenta $k>k_b$. Their ratio,
\[
\frac{L^\ast}{L_c}=\frac{(k_b-k_0)^2\,k_b^2}{2\,\varepsilon_0^4},
\]
is $\simeq 0.17$ for the parameters of the main text, so the above-barrier crossover occurs first. For $L\gtrsim L^\ast$, the transmitted spectrum is no longer described by the sub-barrier saddle: it reduces to the portion of the incident Gaussian above the barrier top, whose effective width is set by the local decay rate of the Gaussian tail at $k_b$,
\[
\sigma_{\rm above}\approx \frac{\sigma_k^2}{k_b-k_0}\ll\sigma_k ,
\]
modulated by the oscillatory (resonant) above-barrier transmission. The temporal spread of this component is therefore parametrically larger,
\[
\Delta\mathcal T_{\rm above}\approx \frac{1}{2\,\sigma_{\rm above}\,v(k_b)}=\frac{k_b-k_0}{2\,\sigma_k^2\,v(k_b)}
\qquad\Big(\simeq 29 \text{ for the parameters of Fig.~2 of the main text}\Big),
\]
exceeding the incident-packet estimate $1/(2\sigma_k v_0)$ by the factor $(k_b-k_0)v_0/(\sigma_k v(k_b))\gg1$. In the crossover region, this broad above-barrier contribution moreover coexists with the residual sub-barrier peak, further inflating the variance. Consequently, the numerically computed spread grows rapidly for $L\gtrsim L^\ast$ (reaching $\Delta\mathcal T_{\rm TF}\simeq 32$ at $L=23$, close to the tail estimate above), and Eq.~(9) of the main text---which tracks only the sub-barrier saddle component---lies below the actual spread in this regime; it should then be regarded only as a lower estimate, no longer providing a controlled approximation to the full distribution. This is illustrated in Fig.~\ref{fig:SMwidths}.}

\subsection{Regional QS distribution}

The regional QS signal is the in-barrier occupation
\begin{equation}
p(t)=\int_0^L |\Psi(x,t)|^2\,dx,
\label{eq:pt_def}
\end{equation}
with
\begin{equation}
\Psi(x,t)=\frac{1}{\sqrt{2\pi}}\int dk\,\phi(k)\,\psi_k(x)\,e^{-iE_k t/\hbar}.
\end{equation}
Substituting into Eq.~\eqref{eq:pt_def} gives
\begin{equation}
p(t)=\frac{1}{2\pi}\iint dk\,dk'\,\phi(k)\phi^*(k')\,Q(k,k';L)\,e^{-i(E_k-E_{k'})t/\hbar},
\label{eq:pt_Qkernel}
\end{equation}
where
\begin{equation}
Q(k,k';L):=\int_0^L dx\,\psi_k(x)\psi_{k'}^*(x).
\label{eq:Qkernel_def}
\end{equation}
Using the Gaussian packet,
\begin{equation}
\phi(k)\phi^*(k')
=
|A|^2
\exp\!\left[-\frac{(k-k_0)^2}{4\sigma_k^2}\right]
\exp\!\left[-\frac{(k'-k_0)^2}{4\sigma_k^2}\right]
e^{-i(k-k')x_0}.
\label{eq:phiphi_star}
\end{equation}

\paragraph{Phase ansatz for the kernel.}
To obtain a simple stationary-phase estimate, we assume that in the narrow-band regime the phase of the overlap kernel is governed by the same barrier phase as the transmitted amplitude, i.e.
\begin{equation}
Q(k,k';L)\approx \mathcal A\!\left(\frac{k+k'}2,q;L\right)
\exp\!\Big(i[\varphi_{\rm TF}(k,L)-\varphi_{\rm TF}(k',L)]\Big),
\label{eq:Q_phase_ansatz}
\end{equation}
with $q=k-k'$. 
This is a heuristic but natural approximation: it captures the leading phase slope responsible for the temporal position of the occupation signal, while its amplitude controls only the normalization and does not affect the leading-order center.

Introducing
\begin{equation}
K=\frac{k+k'}2,
\qquad
q=k-k',
\end{equation}
one has
\[
k=K+\frac q2,
\qquad
k'=K-\frac q2,
\]
and the Gaussian factor becomes
\begin{equation}
\exp\!\left[-\frac{(K-k_0)^2}{2\sigma_k^2}\right]
\exp\!\left[-\frac{q^2}{8\sigma_k^2}\right].
\label{eq:Kq_gaussian}
\end{equation}

\paragraph{Regional QS mean time.}
Expand the phase for small $q$:
\begin{equation}
\varphi_{\rm TF}\!\left(K+\frac q2,L\right)-\varphi_{\rm TF}\!\left(K-\frac q2,L\right)
\approx
q\,\partial_K\varphi_{\rm TF}(K,L).
\label{eq:phase_expand_q}
\end{equation}
Similarly, linearizing the dispersion relation,
\begin{equation}
E_k-E_{k'}\approx \hbar v(K)\,q.
\label{eq:energy_diff_q}
\end{equation}
Using Eqs.~\eqref{eq:phiphi_star}--\eqref{eq:energy_diff_q}, the total phase in Eq.~\eqref{eq:pt_Qkernel} becomes
\begin{equation}
\Phi_{\rm QS}(q,t;K)
=
q\,\partial_K\varphi_{\rm TF}(K,L)-q\,x_0-q\,v(K)t.
\end{equation}
The stationary condition with respect to the relative variable $q$ gives
\begin{equation}
\partial_K\varphi_{\rm TF}(K,L)-x_0-v(K)t=0.
\end{equation}
Since the packet is concentrated near $K=k_0$, we obtain
\begin{equation}
t_{\rm QS}
\approx
\frac{-x_0+\partial_k\varphi_{\rm TF}(k_0,L)}{v_0}
=
-\frac{x_0}{v_0}+\tau_W(k_0,L),
\label{eq:tQS_stationary}
\end{equation}
where
\[
v_0=\frac{\hbar k_0}{m}.
\]

\paragraph{Regional QS spread.}
The Gaussian factor in $q$ in Eq.~\eqref{eq:Kq_gaussian} yields, after integration over the relative momentum,
\begin{equation}
p(t)\propto \exp\!\Big[-2\sigma_k^2v_0^2(t-t_{\rm QS})^2\Big].
\label{eq:pt_gaussian_final}
\end{equation}
Thus the stationary-phase estimate for the regional QS width is
\begin{equation}
\boxed{
\Delta \mathcal T_{\rm QS}^{\rm(reg)}
\approx
\frac{1}{2\sigma_k v_0}.
}
\label{eq:QS_naive_width}
\end{equation}

\paragraph{Opaque-limit saturation of QS.}
Unlike TF, the regional QS signal involves the in-barrier overlap kernel~\eqref{eq:Qkernel_def}. 
In the opaque regime, the exponential suppression associated with $t(k,L)$ is compensated by the in-barrier weight entering $Q(k,k';L)$, so the effective spectral width remains of order $\sigma_k$ rather than shrinking with $L$. 
Accordingly, both the center~\eqref{eq:tQS_stationary} and the spread~\eqref{eq:QS_naive_width} saturate as $L$ increases.

\subsection{Summary of the stationary-phase estimates.}

The above derivation shows the common structure:
\begin{equation*}
\langle \mathcal T\rangle
\approx
\text{free-flight entry time}+\text{phase delay} 
\end{equation*}
and
\begin{equation*}
\Delta \mathcal T
\approx
\frac{1}{2\,(\text{effective spectral width})\,(\text{group velocity})}.
\end{equation*}
For the TF distribution, we have:
\begin{equation}
\boxed{
\langle \mathcal T\rangle_{\rm TF}
\approx
-\frac{x_0}{v(k_\ast)}+\tau_W(k_\ast,L),
\qquad
\Delta \mathcal T_{\rm TF}
\approx
\frac{1}{2\,\sigma_{\rm eff}^{\rm(TF)}(L)\,v(k_\ast)},
}
\end{equation}
with $\sigma_{\rm eff}^{\rm(TF)}(L)\approx \sigma_k$ for $\epsilon_0L\ll1$ and {$\sigma_{\rm eff}^{\rm(TF)}(L)\gtrsim\sigma_k$ in the sub-barrier saddle regime $\eps_0 L\gg1$ with $L\lesssim L^\ast$; beyond $L^\ast$ the saddle estimate no longer applies and Eq.~(9) of the main text lies below the numerical spread, serving only as a lower estimate (see above and Fig.~\ref{fig:SMwidths})}. 

For the QS distribution, we found:
\begin{equation}
\boxed{
\langle \mathcal T\rangle_{\rm QS}^{\rm (reg)}
\approx
-\frac{x_0}{v_0}+\tau_W(k_0,L),
\qquad
\Delta \mathcal T_{\rm QS}^{\rm(reg)}
\approx
\frac{1}{2\sigma_k v_0}.
}
\end{equation}
The key qualitative difference is that TF probes transmitted activity at the exit and therefore {depends on the barrier-reshaped transmitted spectrum}, while regional QS probes in-barrier occupation and remains controlled by the incident spectral width.

%% ============================================================
{\section{Numerical illustration of the distributions and their temporal spreads}
\label{sec:numerics}}
%% ============================================================

\begin{figure}[h!]
    \centering
    \includegraphics[width=0.85\linewidth]{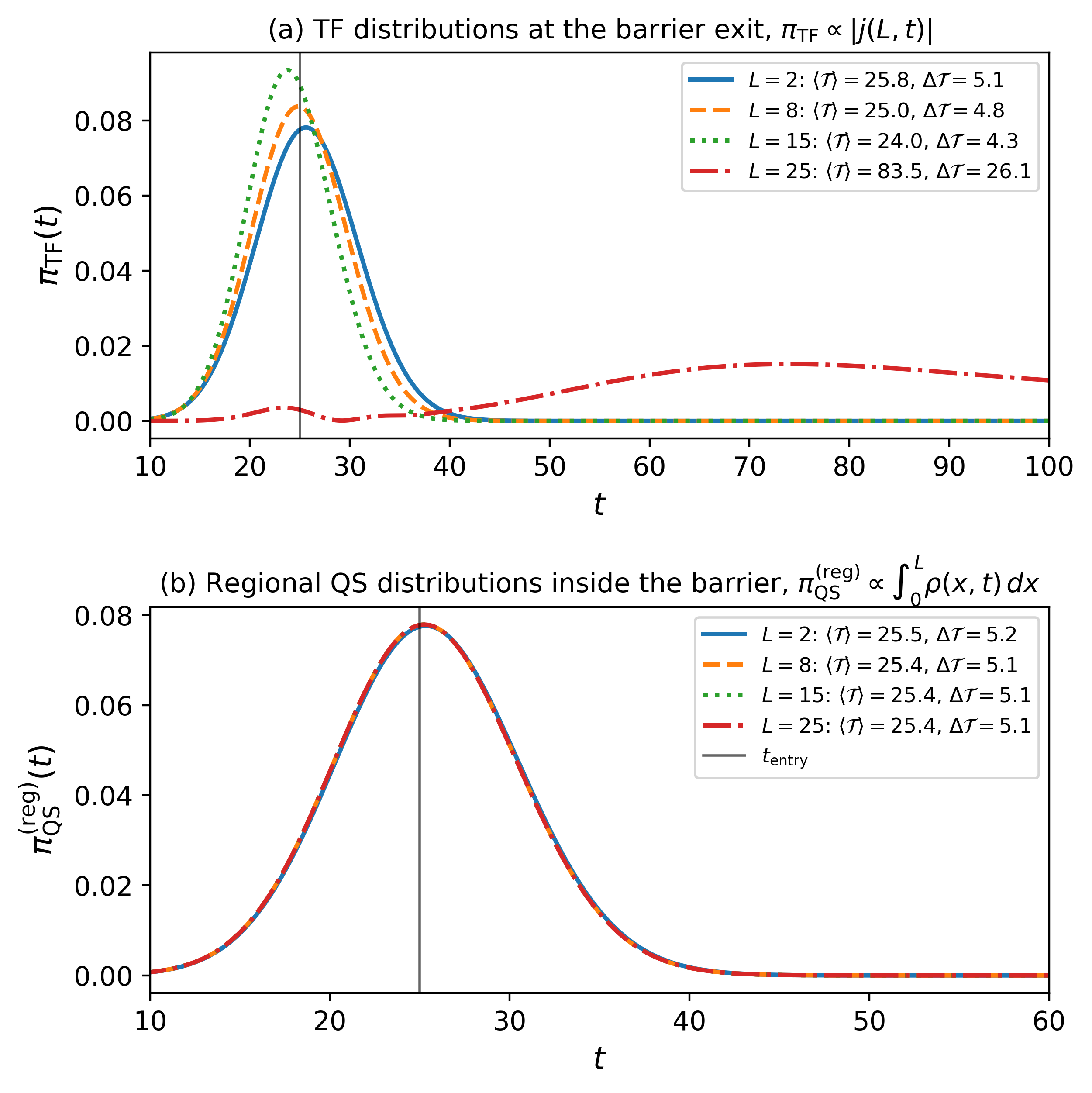}
    \caption{{(a) Normalized TF distributions $\pi_{\rm TF}(t)\propto|j(L,t)|$ at the barrier exit and (b) regional QS distributions $\pi_{\rm QS}^{\rm(reg)}(t)\propto\int_0^L\rho(x,t)dx$, for barrier widths $L=2,8,15,25$. Legends give the numerically computed mean and spread of each distribution; the vertical line marks the classical entry time $t_{\rm entry}=-x_0/v_0$. For $L=25>L^\ast$, the TF distribution develops a broad, resonantly modulated above-barrier contribution, while the QS distributions for $L\ge8$ are essentially superimposed (saturation). Parameters as in Fig.~2 of the main text.}}
    \label{fig:SMdist}
\end{figure}

\begin{figure}[h!]
    \centering
    \includegraphics[width=0.7\linewidth]{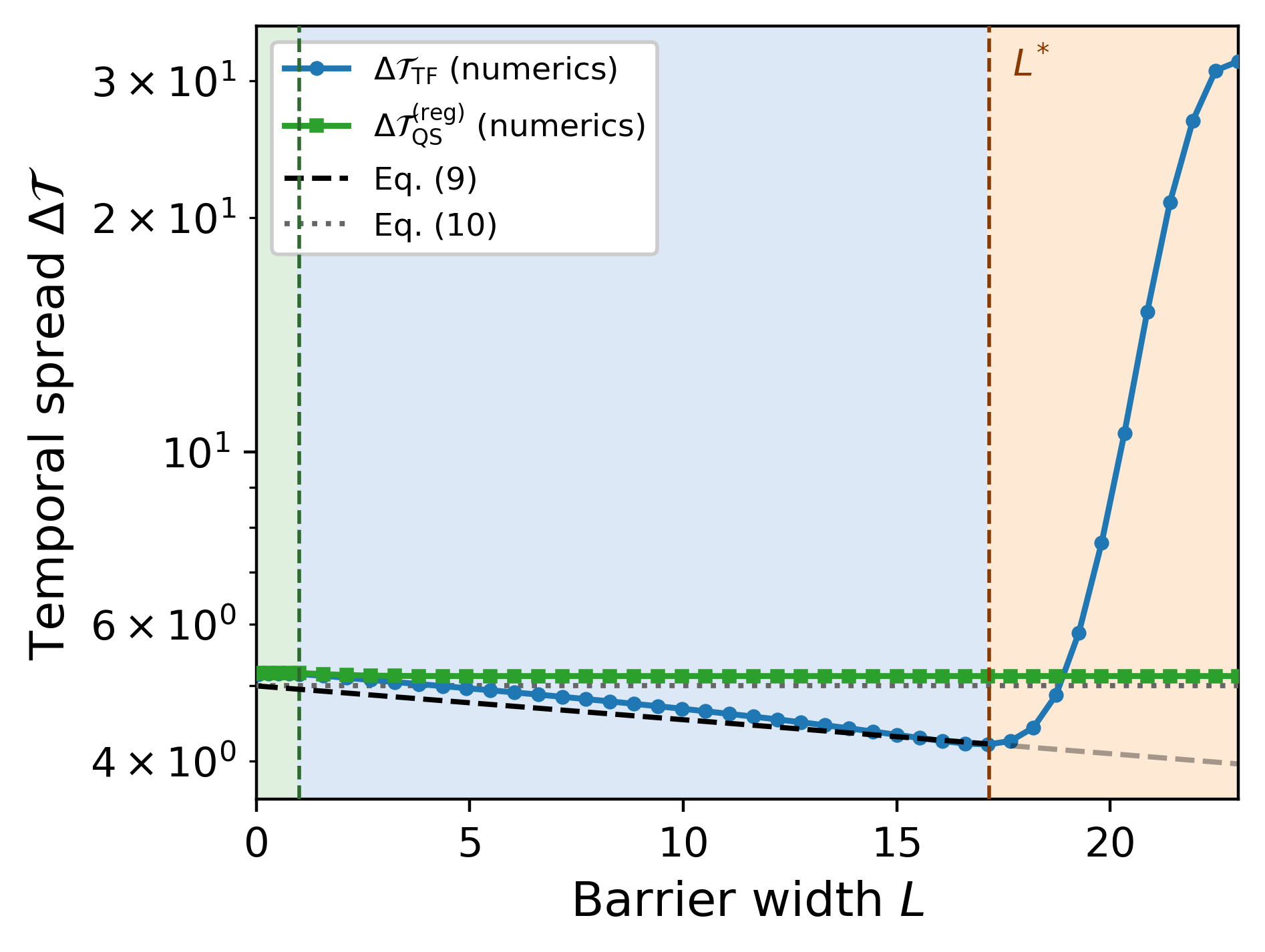}
    \caption{{Temporal spreads $\Delta\mathcal T_{\rm TF}$ (circles) and $\Delta\mathcal T_{\rm QS}^{\rm(reg)}$ (squares) versus barrier width $L${, shown with a logarithmic vertical axis}, compared with the analytical estimates of Eqs.~(9) and (10) of the main text (dashed and dotted lines). Eq.~(9) is tight for $L\lesssim L^\ast$; beyond the above-barrier crossover it lies below the numerical spread, which grows rapidly, and its continuation (faint dashed line) serves only as a lower estimate. The shaded regions match those of Fig.~2 of the main text. Parameters as in Fig.~2 of the main text.}}
    \label{fig:SMwidths}
\end{figure}

{Figure~\ref{fig:SMdist} shows the normalized TF and regional QS distributions for representative barrier widths, computed by direct numerical evaluation of the wave-packet dynamics [Eqs.~\eqref{eq:psi_inside}--\eqref{eq:Tk}] with the parameters of the main text ($m=0.5$, $\hbar=1$, $V_0=2$, $k_0=1$, $\sigma_k=0.05$, $x_0=-50$; momentum grid $k\in[k_0-10\sigma_k,\,k_0+20\sigma_k]$, temporal window $t\in[0,150]$).
For $L=2$, $8$, and $15$ ($L<L^\ast$), the TF distribution is quasi-Gaussian, with a mean that first increases slightly and then advances across the Hartman regime, and a spread that mildly decreases, in agreement with the saddle estimate. For $L=25>L^\ast$, the distribution is strongly non-Gaussian: a residual sub-barrier peak coexists with a broad, resonantly modulated above-barrier contribution extending to much later times. In contrast, the regional QS distributions for $L=8$, $15$, and $25$ are essentially indistinguishable, illustrating the saturation of both their mean and their spread.}

{Figure~\ref{fig:SMwidths} compares the numerically computed spreads with the analytical estimates of the main text. The numerical $\Delta\mathcal T_{\rm TF}$ decreases from $\simeq 5.2$ at $L\to0$ to a minimum $\simeq 4.2$ at $L\simeq L^\ast$, following Eq.~(9) of the main text to within a few percent throughout this range; beyond $L^\ast$ it grows rapidly (reaching $\simeq 32$ at $L=23$, consistent with the above-barrier tail estimate $\Delta\mathcal T_{\rm above}\simeq29$), and Eq.~(9) lies below it, serving only as a lower estimate. The numerical $\Delta\mathcal T_{\rm QS}^{\rm(reg)}$ saturates at $\simeq 5.15$ for $L\gtrsim 3$, within $3\%$ of the $L$-independent estimate of Eq.~(10) of the main text.}

\section*{Entrance current and zero-crossing time}

We analyze the temporal structure of the probability current at the entrance of the barrier, $j(0,t)$, obtained from the same wave-packet scattering setup, see equation \eqref{eq:psi_inside}. For each barrier width $L$, we extract the first time at which the current becomes negative after the main positive peak. This quantity provides a simple dynamical marker for the onset of reflection at the entrance.

\begin{figure}[h!]
    \centering
    \includegraphics[width=1\linewidth]{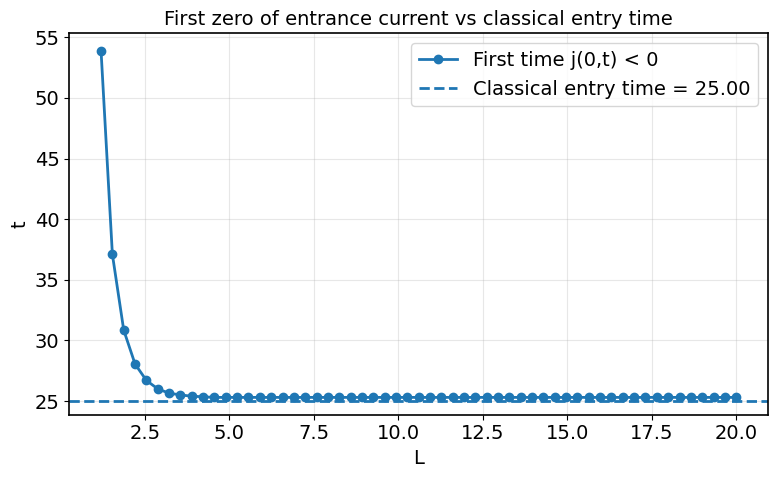}
    \caption{First time at which the entrance current $j(0,t)$ becomes negative as a function of the barrier width $L$. The dashed line indicates the classical entry time $t_{\rm entry} \approx -x_0/v_0$. {For $L\gtrsim4$, the} zero-crossing time remains close to this scale and exhibits only a weak dependence on $L$. (Parameters: $x_0=-50$, $\sigma_k=0.05$, $k_0=1$, $m=0.5$, $\hbar=1$, and $V_0=2$.)}
    \label{Fig1}
\end{figure}

Figure~\ref{Fig1} shows the dependence of this first zero-crossing time on the barrier width $L$. {For thin barriers, transmission dominates and the first sign change occurs only in the late tail of the packet; for $L\gtrsim4$, the zero-crossing time remains close to the classical entry time $t_{\rm entry} \approx -x_0/v_0$ and depends only weakly on $L$.} This supports the statement made in the \emph{discussion} paragraph of the paper that{, in the opaque regime,} the change of sign of the entrance current occurs shortly after the arrival of the incident wave packet at the barrier.

\begin{figure}[h!]
    \centering
    \includegraphics[width=1\linewidth]{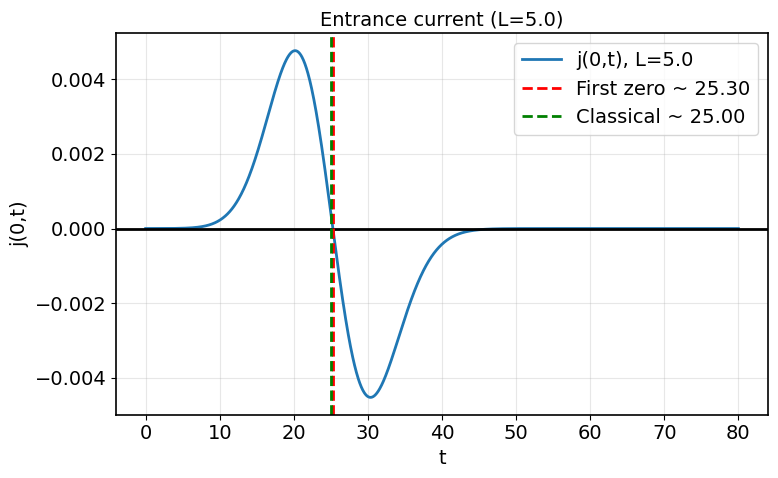}
    \caption{Entrance current $j(0,t)$ as a function of time for a representative barrier width. The current shows a positive peak associated with the incoming wave packet, followed by a sign change due to reflection. {The red dashed vertical line indicates the first zero-crossing time, and the green dashed line the classical entry time $t_{\rm entry}$.} (Parameters: $x_0=-50$, $\sigma_k=0.05$, $k_0=1$, $m=0.5$, $\hbar=1$, and $V_0=2$.)}
    \label{Fig2}
\end{figure}

Figure~\ref{Fig2} illustrates the temporal profile of the entrance current $j(0,t)$ for a representative barrier width. The current exhibits an initial positive peak associated with the incoming wave packet, followed by a sign change due to the reflected component. The first zero-crossing provides a well-defined temporal marker for the onset of reflection.

\end{document}